%% file: main.tex
\documentclass[10pt]{article}

\usepackage[a4paper]{geometry}
\usepackage[T1]{fontenc}
\usepackage[utf8]{inputenc}
\usepackage{amsmath,amssymb}
\usepackage{graphicx}
\usepackage{booktabs}
\usepackage{float}
\usepackage{natbib}
\usepackage{siunitx}
\usepackage{hyperref}
\usepackage{footmisc}

\title{\huge Observational daily and regional photovoltaic solar energy production for the Netherlands}
\author{B. P. M. Laevens\footnote{bpm.laevens@cbs.nl}$~^{1,2}$, O. ten Bosch$^1$, F. P. Pijpers$^1$ and W.G.J.H.M. van Sark$^3$}
\date{%
    $^1$\textit{Statistics Netherlands, The Hague, The Netherlands}\\%
    $^2$\textit{Ministry of Economic Affairs and Climate Policy, The Hague, The Netherlands}\\
    $^3$\textit{Copernicus Institute of Sustainable Development, Utrecht University, Utrecht, The Netherlands}\\[2ex]%
    \today
}

\begin{document}

\maketitle

\input{abstract.tex}
\clearpage
\input{s1_introduction.tex}
\input{s2_solar_background.tex}

\input{s3_data.tex}

\input{s4_method.tex}

\input{s5_representativeness}

\input{s6_results.tex}

\input{s7_summary.tex}
\input{acknowledgements.tex}

\bibliography{main.bib}

\end{document}

%% file: abstract.tex
\section*{Abstract}
This paper presents a classical estimation problem for calculating the energy generated by photovoltaic solar energy systems in the Netherlands, on a daily, annual and regional basis. We identify two data sources to construct our methodology: pvoutput, an online portal with solar energy yield measurements, and modelled irradiance data, from the Royal Netherlands Meteorological Institute. Combining these, we obtain probability functions of observing energy yields, given the irradiance, which we then apply to our PV systems database, allowing us to calculate daily and annual solar energy yields. We examine the variation in our daily and annual estimates as a result of taking different subsets of pvoutput systems with certain specifications such as orientation, tilt and inverter to PV capacity ratio. Hence we obtain specific annual energy yields in the range of 877--946 $kWh/kWp$ and 838--899 $kWh/kWp$ for 2016 and 2017 respectively. The current method used at Statistics Netherlands assumes this to be 875 $kWh/kWp$, meaning the yields were underestimated and overestimated for 2016 and 2017 respectively. Finally, we translate our national estimates into solar energy yields per municipality. This research demonstrates that an irradiance based measure of solar energy generation is necessary to obtain more accurate energy yields on both a national and regional level.

%% file: s1_introduction.tex
\section{Introduction}
At the 2015 Paris climate agreement, part of the \textit{United Nations Framework Convention on Climate Change}, nations committed themselves to limiting temperature rise to $1.5^\circ$C, relative to pre-industrial levels\citep{UN2019}. Around the world, nations are devising policies to transition away from polluting fossil fuels towards clean and renewable energy sources. A recent update, presented by the European Commission, as part of the European Green Deal policy, set a target of 55\% greenhouse gas emission reduction by 2030 \citep{EU2030target,EUgreendeal}. In 2018, the share in renewables in the Netherlands stood at 7.2 \% \citep{euobserver}, with the earlier EU 2020 target of 14\% probably not fulfilled \citep{EU2017}. Statistics Netherlands (SN) is the body responsible for publishing renewable energy statistics in the Netherlands \citep{SNstatline}. With policy measures, targeted at increasing the renewables share, starting to take effect, attention to detailed and accurate measurements of renewable energy yields will become ever more important.\\

\noindent The Center for Big Data Statistics at SN was established in 2016 to examine and develop new methods for official statistics based on new and alternative data sources. In this paper, we report on a new method to estimate regional and national, daily and annual solar energy yield for the Netherlands. The emphasis of our new method lies in using and combining readily available measured energy yield data from photovoltaic sytems in the Netherlands as well as meteorological data, allowing us to generate, for the first time, a weather based energy yield estimate. In section 2 we provide the reader with an introduction to the field of solar energy, followed by a description of the new data sources (and cleaning) that are used in our method in section 3. We outline in detail our methodology in section 4, paying particular attention to aspects of representativeness in section 5. Our results are discussed in section 6, before oncluding in section 7.

%% file: s2_solar_background.tex
\section{Solar Energy}
In section 2.1, we introduce the current solar yield method at SN as well as a few concepts relating to solar energy. This will also enable us to contrast the current method with our newly proposed one, which we outline in section 2.2.

\subsection{Current method}
\noindent Photovoltaic (PV) systems in the Netherlands have seen an explosive growth over the past decade. In just ten years the installed capacity has risen from 0.09 Gigawatts (GW) in 2010 to 6.9 GW in 2019 \citep{SNstatline}. Most of the systems are recorded in a PV systems database, containing exact location ($L$), the system's installation date ($d$) and its power or systemsize ($P$). In the past, small households made up the bulk of the PV systems, but as of 2019 just over half of the capacity is made up of large PV systems ($>15~kWp$) such as solar parks \citep{SNlargevssmall}. Subsidy schemes for large PV systems have meant that their energy yields are recorded on a monthly and hence annual basis \citep{CertiQ}. The energy yields of small households, by contrast, are not collected nationally. \\

\noindent A framework for measuring the total energy yield from household PV systems was introduced to combat the paucity of measurement data and is outlined in the Netherlands Enterprise Agency's protocol of renewable energy \citep{RVO2019}. At the centre of this framework is an equation (\ref{equE}) that translates the average installed PV capacity in a specific year, expressed in Watts ($W$) or Watts--peak ($W_{p}$), into an average annual energy yield ($Y_{a}$), expressed in units of energy e.g. Watt hours ($Wh$), through means of a representative constant, denoted as the specific annual yield ($Y_{s,a}$). With Watt--peaks we mean the number of Watts produced at peak power. 

\begin{equation}
Y_{a} = \left(\frac{\sum\limits_{n=1}^{N_{1}}P_{n} + \sum\limits_{n=1}^{N_{365}} P_{n}}{2}\right)~Y_{s,a}
\label{equE}
\end{equation}

\noindent where $P_{n}$ is the system size of one PV system in the database, $N_{1}$ and $N_{365}$ the number of PV systems in the database on the first and last days of the year respectively. Finally $Y_{s,a}$ is the specific annual yield, currently fixed at 875 $kWh/kWp$ \citep{Sark2014}.\\ 

\noindent Two datasets of PV systems' performances from 2012 and 2013 were used to determine $Y_{s,a}$, yielding $877 \pm 140$ and $874 \pm 140$ $kWh/kWp$ for both years in question\citep{Sark2014}. While this research pinpointed a value for $Y_{s,a}$ in those years, it could be improved upon in terms of corrections for lack of representativeness. It is known from the literature that various aspects of PV systems are influential in determining $Y_{s,a}$ or $Y_{a}$ \citep{eia}. These are variables such as orientation ($\phi$), tilt ($\theta$), inverter size ($P_{i}$), irradiance $H$, longitude ($l$) and latitude ($b$) and technological advancement\\

\noindent The orientation angle can be expressed in cardinal signs or degree angles ranging from \ang{0} to \ang{360} (with \ang{180} conventionally meaning South--oriented in the Northern hemisphere) whereas $\theta$ ranges from \ang{0} (flat, facing up to the sky) to \ang{90} (facing towards the horizon). PV systems comprise of PV panels and one or more inverters, whose main function is to convert the DC, generated by the PV systems, into AC, that is fed back into the grid. It is quite customary for inverters to be undersized, i.e. $P_{i}<P$ \citep{burger2006,tsafarakis2017}. One reason for doing this could be that a system's $\theta$ and $\phi$ and hence $Y$ are suboptimal and will never reach the full potential of $P$, making a smaller $P_{i}$ sufficient. A second reason for a smaller $P_{i}$ is that it can optimise the system's efficiency in low--light conditions e.g. mornings, evening and cloudy days at the expense of losing (or clipping) some $Y$ at the peak performance time, e.g. midday \citep{smaoversizing, solaredgeoversizing}. Though there is no benefit in terms of augmenting $Y$, oversizing ($P_{i}>P$) also occurs, by owners who may wish to increase $P$ in the future \citep{solarchoice}. When extrapolating $Y_{s,a}$ to the whole country, it is important to account for the true distributions of $\phi$, $\theta$ and $\epsilon=\frac{P_{i}}{P}$ (the inverter to PV capacity ratio).\\

\noindent Finally, it is important to account for the geographical distribution (longitude $(l)$ and latitude $(b)$) of the systems since weather varies locally. The most important weather factor is the irradiance ($H$), which is a measure of the amount of solar power per unit area that is incident on Earth, expressed in units of $W/m^{2}$ or energy per unit area ($Wh/m^2$). It is known that Western parts of the Netherlands receive up to 10\% more $H$, on an annual basis, compared to Eastern parts \citep{litjens2017}. Western parts of the country have the highest population density, implying high density of household PV systems, whereas sparser Northern and Eastern areas allow for larger PV systems such as solar parks.

\begin{figure}[!htb]
\centering
\includegraphics[width=0.9\linewidth]{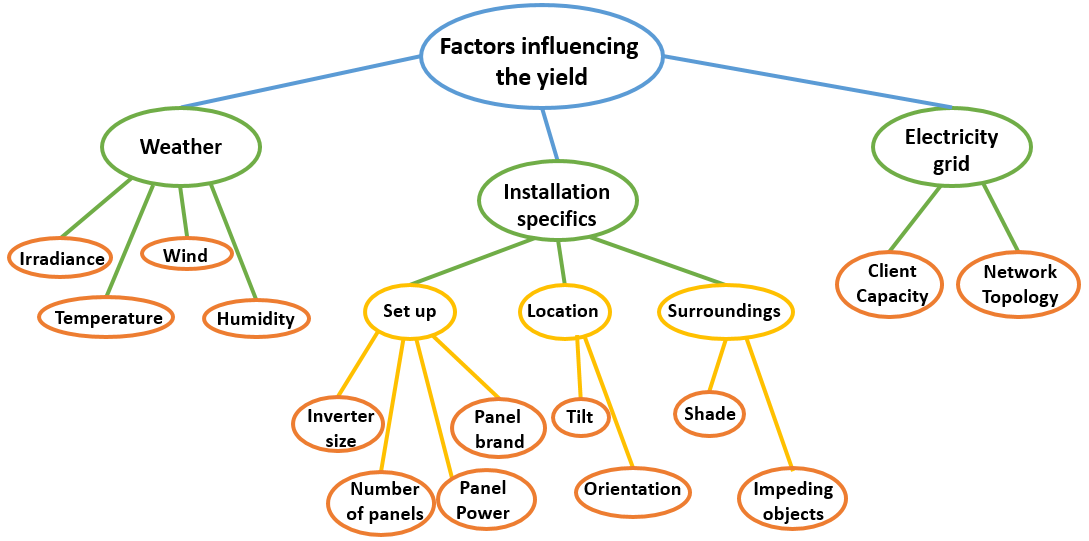}
\caption[Different factors that can influence PV yields]{An overview of different factors that can influence the performance of PV systems.}
\label{factors}
\end{figure}

\noindent A disadvantage of using equ. \ref{equE} is the assumption that $Y_{s,a}=875~kWh/kWp$, regardless of the year in question. The total annual irradiance ($H_{a}$) equalled 989 and 1003 $kWh/m^2$ in 2012 and 2013 respectively, which is similar to the 30 year average of 986 $kWh/m^2$. The past four years (2016--2019) have been substantially sunnier with $H_{a}$ equalling 1040, 1020, 1137 and 1099 $kWh/m^2$ \citep{knmi2016_jow, knmi2017_jow, knmi2018_jow, knmi2019_jow}. It therefore follows that $Y_{s,a}$ should in fact be higher than $875kWh/kWp$ for these years. In Figure \ref{factors} we show a schematic overview of different factors that can influence $Y$, some of which we discussed in the paragraphs above. For more information regarding photovoltaics, please consult \citet{reinders}.

\subsection{New method}
In light of the discussion we have set out in the previous section, we have developed a new method to estimate the solar energy yield on a daily ($Y_{d}$) and annual ($Y_{a}$) basis, which we translate into regional estimates. Our aims are two--fold: we wish to assess how accurate the current SN method is for estimating $Y_{a}$, when taking into account PV system characteristics such as $\phi$, $\theta$, $\epsilon$, $l$, $b$ and incident $H$. Secondly, we wish to provide high resolution granular information as a useful tool for local authorities when devising their energy transition policies.\\

\noindent Our new method is a classical statistical estimation problem: a population of PV systems exists for which we wish to estimate $Y$, but we only have a non--probability sample of measurements which we need to extrapolate to the whole population. We have three data sources that allow us to do this:
\begin{enumerate}
\item pvoutput \citep{pvoutput_latest}, an online portal, containing $Y$ measurements of PV systems, along with metadata such as: $\phi$, $\theta$, $P$, $P_{i}$ and postal code amongst others.
\item irradiance data from the Royal Netherlands Meteorological Institue \citep{knmi_about}. Irradiance data is available every 15 mins on a grid resolution of 3x6 $km^2$.
\item PV systems database, containing almost all PV systems in the Netherlands. For each system we know: $L$, $P$ and $d$.
\end{enumerate}

\noindent By linking sources (a) and (b), through $L$, we infer relations between $H_{d}$ and $Y_{d}$ for our non--probability sample. And by combining sources (b) and (c), again through $L$, we obtain $H_{d}$ distributions for the PV systems database. The information from (a) and (b), i.e. $Y_{d}$ given $H_{d}$ is then used to estimate $Y_{d}$ for the whole population. We investigate the sensitivity of our results for $Y_{d}$ by selecting different priors relating to different distributions of PV systems characteristics ($\phi$, $\theta$ and $\epsilon$).

%% file: s3_data.tex
\section{Data}
\subsection{Data Sources}
\subsubsection{pvoutput}
Pvoutput is an Australian online portal with near real--time information of PV energy generation at various locations throughout the world. After Australia, the Netherlands has the largest share (20.57\%) of installed capacity, registered on the website. As of October 2019, this capacity stands just shy of 50 MW \citep{pvoutput_latest}, which is around 0.9\% of the total capacity in the Netherlands. The number of Dutch PV systems in 2016 and 2017 was on the order of 5600. Other countries with substantial registered capacity are the USA, Italy, Germany, the UK and Belgium. The metadata contain the following specifications:

\begin{enumerate}
\item power: Number of panels ($N_{p}$), panel power ($P_{p}$), total systemsize ($P$), inverter size ($P_{i}$) and number of inverters ($N_{i}$).
\item geometry: orientation ($\phi$) and tilt ($\theta$).
\item brand: panel and inverter brands.
\item location: Postal code 4 (pc4) area, longitude ($l$) and latitude ($b$).
\item time: installation date of the system ($d$).
\item comments: explanatory comments can be added by the user.
\end{enumerate}

\noindent For more details regarding this metadata, please see \citet{pvoutput_overview}. The data itself consist of the following variables:

\begin{enumerate}
\item energy: instantaneous ($Y_{inst}$) and cumulative energy measurements ($Y_{cum}$) 
\item time: date and time.
\end{enumerate}

\subsubsection{Dutch meteorological weather data}
The \textit{Koninklijk Nederlands Meteorologisch Instituut} or Royal Dutch Meteorological Institute (hereafter KNMI) is the body responsible for measuring variables related to the weather in the Netherlands \citep{knmi_about}. Besides the 30 different ground--based weather stations, which measure different variables such as wind speed, rainfall, temperature and irradiance, the institute has also developed a physics based empirical model to calculate many of these same variables \citep{deneke2008, greuell2013, knmiviewer}. This model uses input data from the \textit{Spinning Enhanced Visible and Infrared Imager} (SEVIRI) instrument \citep{seviri} on board the \textit{Meteosat Second Generation} satellites, located in a geostationary orbit at 36.000 km. These are operated by EUMETSAT \citep{eumetsat}. SEVIRI observes properties of the atmosphere every 15 minutes and has a resolution of 3x3 $km^2$ at nadir \citep{knmisatellite}. Due to projection effects, this results in a resolution of 3x6 $km^2$ for the Netherlands. It is important to note that the irradiance data are only available at times when the Sun's elevation exceeds $12^\circ$ \citep{greuell2013}. The modelling does not work well outside this regime due to 3D cloud effects. These data contain the following variables:

\begin{enumerate}
\item irradiance ($H$) in units of $kW/m^2$
\item grid cell centre $l$ and $b$
\end{enumerate}

\subsubsection{PV Systems Database}
Statistics Netherlands constructs its own database of PV systems based on different data sources. The biggest and most important of these is the \textit{Product Installation Register} or PIR \citep{SNsolar}, provided by the network operators. This data source is incomplete and is estimated to contain around $\sim 85\%$ of small household systems. This is why, in recent years, additional systems, not present in the PIR, have been identified using data from the Dutch tax authority. People are incentivised to purchase PV systems by registering a VAT return for the cost and installation of the panels. Although this is not obligatory, it is assumed that this can entice a lot of people to do so. Even after adding tax returns the register is still incomplete. This data source contains the following variables:

\begin{enumerate}
\item power: total systemsize ($P$).
\item time: installation date ($d$).
\item location: house number and postal code.
\end{enumerate}

\noindent We have a far more complete view on large PV systems in the Netherlands. Statistics Netherlands has administrative data on $P$ and monthly yield ($Y_{m}$) of such solar farms from the government--led certification process for renewable energy as executed by CertiQ \citep{CertiQ}, a 100\% subsidiary of TenneT, the European electricity transmission system operator for the Netherlands \citep{TenneT}. We do not use the CertiQ data because our model is aimed at computing $Y_{d}$ (rather than $Y_{m}$), as a function of different PV system parameters such as $\phi$, $\theta$ and $\epsilon$, all of which we also do not know for the CertiQ systems. We do, however, use the CertiQ data to compare with our model, in Section 6. Finally, a project is under way at SN to use machine learning techniques to identify missing solar panels from aerial images (e.g. section 3.2.1 in \citet{debroe2019}).

\subsection{Data Cleaning and Processing}
\subsubsection{pvoutput}
We identified two data cleaning challenges for pvoutput. The first was uniformising and correcting the metadata such that apparently similar quantities, inputted by different owners, mean the same thing. The second involved analysing the real--time pvoutput data to determine how realistic the data is, also in relation to the metadata. 

\paragraph{Uniformising and correcting the metadata}
\noindent While most variables are reasonably consistent (e.g. $\phi$ is limited to choosing a cardinal sign), the reliability of some other variables is less clear e.g. $P_{p}$, $N_{p}$ and $P$. If inputted correctly, then $P=N_{p}\cdot P_{p}$; however, for about 10\% of all PV systems, this was not the case. To resolve this, we chose $P$ to be the more reliable variable, on the assumption that pvoutput owners were more likely to make a mistake concering $N_{p}$ or $P_{p}$.\\

\begin{figure}[!htb]
\centering
\includegraphics[width=0.92\linewidth]{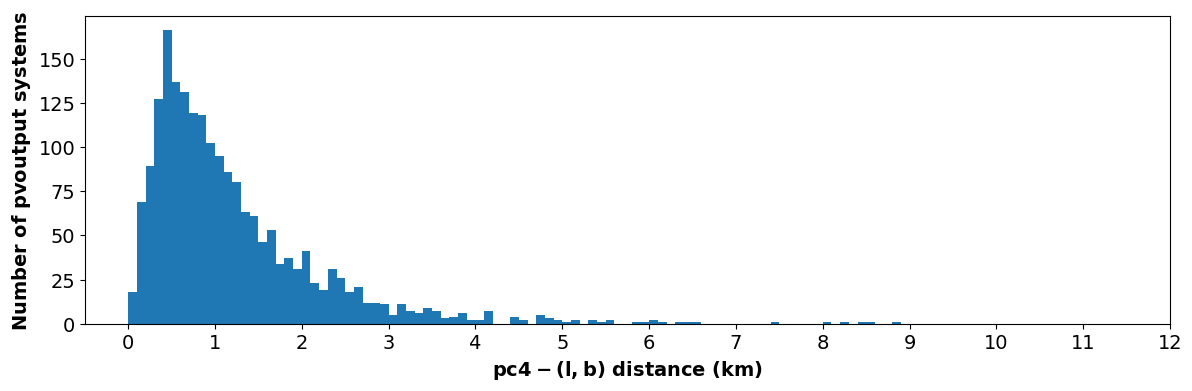}
\caption[Histogram of lat/lon--postal code differences for reliable PV systems in 2016]{PC4 centroid--lat/lon distances for the reliable set of PV systems in 2016.} 
\label{Distance_LonLat_PC4}
\end{figure}

\noindent Since pvoutput contains two types of co--ordinates: pc4 and $(l,b)$, we check the reliability of both of these by comparing them to each other. Postal codes in the Netherlands consist of four digits (pc4), followed by two letters. The pc4 provided in pvoutput, narrows the locations down to a spatial area of $\sim 1-8~km^2$. The $(l,b)$ allow for pinpointing an exact location. For each pvoutput location, we calculated the distances between $(l,b)$ and the pc4 centroids. For about 5\% of these systems the distance is more than 5 km. Manual inspection of the pc4 -- $(l,b)$ distances for a few municipalities shows an interesting pattern: half of the households took the effort to indeed specify $(l,b)$ nearby their dwelling, whereas the other half seems to have chosen a generic $(l,b)$ based on the municipality. These findings led us to retain the pc4 centroid as the location at the expense of losing -- in some cases -- more accurate information from $(l,b)$, where we assume that owners are more likely to know what their pc4 is. This level of precision is also sufficient for our research goals. Figure \ref{Distance_LonLat_PC4} displays the pc4 -- $(l,b)$ distances for the reliable set (see next section) of systems in 2016.

\begin{figure}[!htb]
\centering
\includegraphics[width=0.92\linewidth]{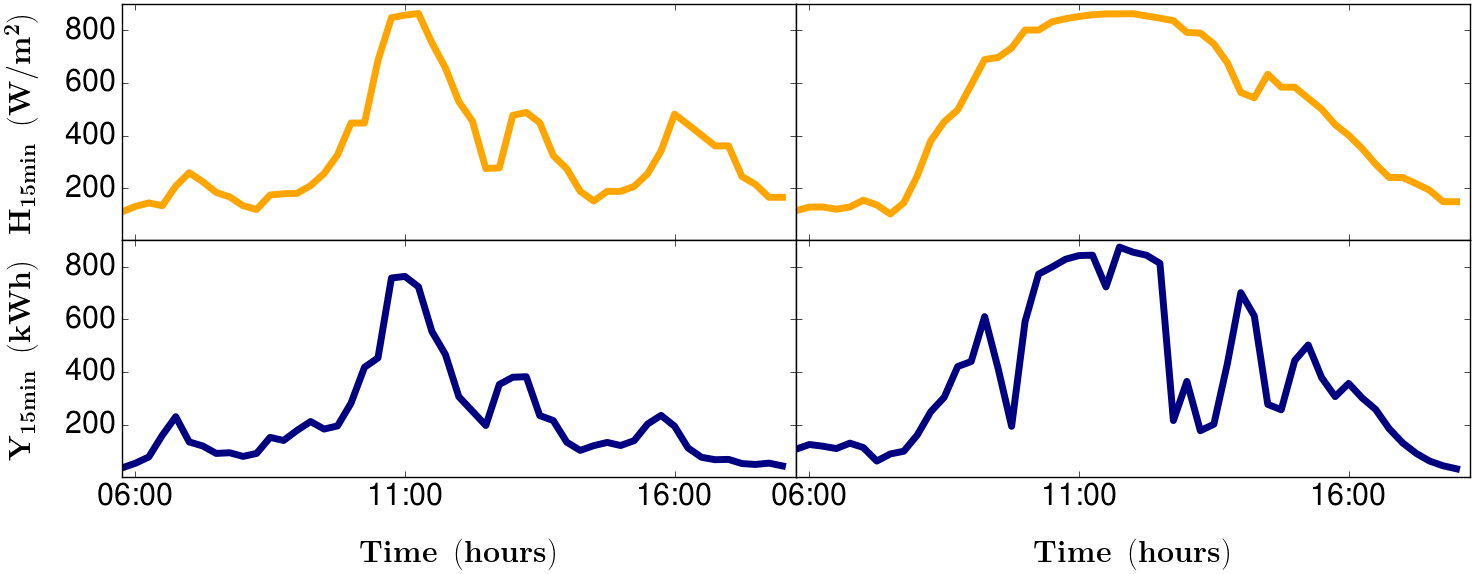}
\caption{\label{pvoutput_knmi}$H$ (orange) and $Y$ (blue) at two different locations on 19/05/2016.}
\end{figure}

\paragraph{Constructing a reliable set of measurements per day}
Having corrected the metadata, we now turn to the measurements. Two examples of pvoutput $Y$-profiles can be seen in Figure \ref{pvoutput_knmi} (blue lines). We devise four different quality criteria, which are performed per PV system and day. Daily measurements are deemed reliable if they pass all four checks. \\

\noindent In Section 3.1.1 we saw that two different energy measurements are provided by pvoutput: $Y_{inst}$ and $Y_{cum}$. The reliability of the latter can be checked by re--calculating the cumulative measurements ($Y_{cum, calc.}$), using the former (equ. \ref{equcumcalc}). A measurement is deemed acceptable if it satisfies equ. \ref{equleeway}, i.e. allowing for a 10\% relative difference with $Y_{cum, calc}$. The second check involves identifying the peak energy per day of a system ($P_{inst, peak}$) and checking it makes sense when comparing to $P$ quoted in the metadata (equ. \ref{equpeak}), allowing for a value exceeding $P$ by 20\%. The second criterion could be too harsh on some days when cloud--induced superirradiance is capable of temporarily producing $P_{inst, peak}$ which is higher than P \citep{zhang2018}. We decide not to take this into account as it is difficult for us to ascertain when this local effect may be occurring. 
 
\begin{gather}
Y_{cum,calc.} = \sum_{n=1}^{N}Y_{inst}\Delta t
\label{equcumcalc}\\
0.9~Y_{cum}<Y_{cum,calc.}<1.1~Y_{cum}
\label{equleeway}\\
P_{inst,peak}\leq 1.2P
\label{equpeak}
\end{gather}

\noindent The third quality check (equ. \ref{equdeltat}) examines the time intervals ($\Delta t$) between each $Y_{inst}$ on a given day. Some PV systems suffer from measuring gaps e.g. a system may measure $Y_{inst}$ consistently, during several hours, with $\Delta t=5~m$, before recording a gap e.g. $\Delta t = 2~h$. This check, in turn, influences quality check number one (equ. \ref{equcumcalc}). The final check sees if there is a measurement for each day of the year for a given PV system (equ. \ref{equmeas}). 

\begin{gather}
\Delta t\leq 15~mins
\label{equdeltat}\\
N_{meas}>0
\label{equmeas}
\end{gather}

\noindent It should be noted that the decision to remove PV systems for certain days resulting from equ. \ref{equdeltat} and \ref{equmeas} is not straightforward. We cannot ascertain whether the gaps or missing days are as a result of Wi--Fi stability issues and/or malfunctioning software or whether the PV system is really not producing any energy or has been turned off. The blue line in Figure \ref{Nids_per_day_comparison} shows the number of systems per day for 2016 after data cleaning is performed. 

\subsubsection{Dutch meteorological weather data}
\begin{figure}[!htb]
\centering
\includegraphics[width=0.8\linewidth]{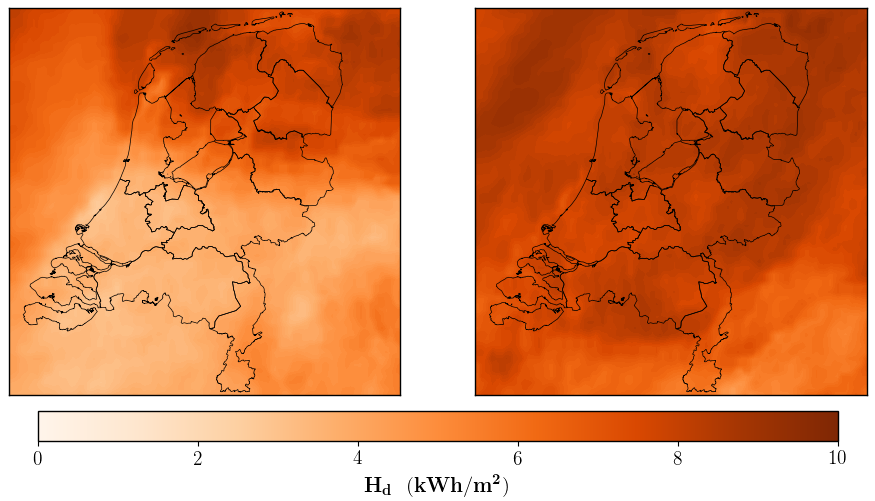}
\caption[Map of total daily irradiance in the Netherlands]{$H_{d}$ per surface area ($kWh/m^2$) for two different days in June: 21 (left) and 22 (right).}
\label{irradiance}
\end{figure}

We do not perform data cleaning on the KNMI irradiance data, since these have been thoroughly checked by KNMI itself. Having downloaded the quarterly hour $H_{k}$ data, we aggregate $H_{k}$ to daily totals $H_{d}$ (equ. \ref{equirr}), following the same procedure as equ. \ref{equcumcalc}.

\begin{equation}
H_{d}=\sum_{k=1}^{K}H_{k} \Delta t
\label{equirr}
\end{equation}

\noindent where $\Delta t=15~mins$. We notice that, occasionally, an $H_{k}$ is missing and we therefore adapt $\Delta t$ in equ. \ref{equirr} accordingly. We decided not to compare modelled $H$ data with weather station $H$ data because this has already been done in \citep{greuell2013}, who showed that discrepancies between the two are minimal and negligible. Figure \ref{irradiance} shows $H_{d}$ for two consecutive days in June 2016, nicely illustrating the effect of different weather conditions.

%% file: s4_method.tex
\section{Methods}
We outline our new method for determining $Y_{d}$ and $Y_{a}$ in the Netherlands. An overview, in formal notation is presented in section 4.1. The different aspects of the method are expanded upon in sections 4.2--4.4. 

\subsection{Procedure}
Our main aim is to calculate $p_{d}(Y)$: the distribution of $Y_{d}$ emanating from population of PV systems in the Netherlands, which we can aggregate to $Y_{a}$. It is therefore trivial to state that $Y_{d}$ and $Y_{a}$ are defined by equ. \ref{equEday} and \ref{equEyear}:

\begin{gather}
Y_{d}=\int p_{d}(Y)dY
\label{equEday}\\
Y_{a}=\sum_{d=1}^{d=365}Y_{d}
\label{equEyear}
\end{gather}

\noindent We stated in Section 2.2, that $Y$ is a function of the weather and more specifically $H$. Therefore $p_{d}(Y)$ can be re--expressed in terms of $H$, given by equ. \ref{equpE1}. De--constructing $p_{d}(Y,H)$ into two separate functions $p_{d}(Y|H)$ and $p_{d}(H)$ and integrating this over the number of PV systems in the database ($N_{d}$) gives equ. \ref{equpE2}:

\begin{gather}
p_{d}(Y) = \int p_{d}(Y,H)dH
\label{equpE1}\\
p_{d}(Y) = \int^{N_{d}} p_{d}(Y|H)p_{d}(H)dH = \int^{N_{d}} p_{d}(Y_{s}|H)p_{d}(P)p_{d}(H)dH
\label{equpE2}
\end{gather}

\noindent Where in the final step, we have re--expressed $p_{d}(Y)$ in terms of $p_{d}(Y_{s})$ and $p_{d}(P)$, the specific yield and power of the systems respectively. Evaluating $p_{d}(H)$ is trivial: the distribution of $H$ of all database locations is obtained by matching the database systems to the nearest grid cell. Evaluating $p_{d}(Y_{s}|H)$ is more involved since we construct this with our non--probability sample pvoutput, which we use as a proxy for the population. We can evaluate $p_{d}(Y_{s}|H)$ as the marginal likelihood, in terms of the PV system characteristics $x=\{\phi,~\theta,~\epsilon\}$, given by equ. \ref{equmarg}:

\begin{equation}
p_{d}(Y_{s}|H) = \int p_{d}(Y_{s}|x)p_{d}(x|H)dx
\label{equmarg}
\end{equation}

\noindent This 3D integral can be approximated using Monte Carlo sampling. We can repeatedly draw samples for $p_{d}(Y_{s}|x)$ from pvoutput that satisfy criteria defined by our choice for the prior $p_{d}(x|H)$, thus obtaining different `realisations' of the pvoutput data, linking $H_{d}$ and $Y_{d}$. We can randomly draw one of the $p_{d}(Y_{s}|H)$ and insert this into equ. \ref{equpE2}. Evaluating this equation repeatedly and hence also equ. \ref{equEday} allows us to build a probability density function for $Y_{d}$, enabling us, in turn, to estimate the mean energy $\mu_{Y_{d}}$ and the standard deviation $\sigma_{Y_{d}}$. \\

\noindent Our calculation for equ. \ref{equmarg} will strongly depend on our choice of prior relating to the distribution of $x=\{\phi,~\theta,~\epsilon\}$. By exploring various scenarios, i.e. different choices of our prior, we can explore the margins of our estimates. We will expand these ideas and equations in sections 4.2--4.4 and 5 and discuss our results in section 6.1.\\

\noindent Finally, we can return to the database systems and determine their yield more accurately. We make the assumption that $\mu_{Y_{s,d}}$ must correspond to $\mu_{H_{d}}$. Therefore, for a system at a location j, we can read off once again $H_{d,j}$ and calculate its offset relative to $\mu_{H_{d}}$ and offset $Y_{s,d\ j}$ accordingly (equ. \ref{equbetajd}). It is then possible to aggregate each system to a regional level of our choosing such as a municipality, given by equ. \ref{equbetamun}. We discuss our regional results in section 6.2.

\begin{gather}
Y_{s,d\ j} = \mu_{Y_{s,d}} + \mu_{Y_{s,d}}\left(\frac{H_{d,j}}{\mu_{H_{d}}}-1\right)
\label{equbetajd}\\
Y_{d,mun} = \sum\limits_{j=1}^{N_{d, mun}}Y_{s,d\ j}P_{j}
\label{equbetamun}
\end{gather}


\subsection{Combining irradiance with pvoutput and the PV systems database}
In section 4.1 we saw that our procedure relies on coupling $H_{d}$ twice: once with pvoutput and another time with our PV systems database. The former being necessary so we can determine $p_{d}(Y_{s}|H)$, whereas the latter enables us to evaluate $p_{d}(H)$. In either case, we approximate all addresses within a pc4 area to the centroid of that area in geographic co--ordinates $(l,b)_{k}$ (equ. \ref{equpc4lb}). Using the Haversine formula \citep{mendoza1797}, the $(l,b)_{k}$ and $(l, b)_{j}$ (cell centroids of $I$) distances may be calculated, thus identifying the closest cell (equ. \ref{equhaversine}).

\begin{equation}
pc4_{k} \sim (l,b)_{k}
\label{equpc4lb}
\end{equation}

\noindent 
\begin{equation}
d_{k}=\min\limits_{j=1}^{N_{cells}}{|(l,b)_{k}-(l,b)_{j}|,~~\forall k \in [1,N_{pc4}]}
\label{equhaversine}
\end{equation}

\noindent This approach makes two assumptions. Firstly, the assignment to the nearest grid cell is correct. Secondly, weather behaves according to the resolution size of a grid cell $H$. The first will not always hold since pc4 areas can range from $\sim$ 1 to 8 $km^{2}$ in size \citep{nemo}, because these are effectively a proxy for population density. Since $H$ grid cells are 3x6 $km^2$ in size, it is conceivable that locations in larger pc4 areas are not always coupled correctly. The second assumption will not always hold because weather can be more local than the resolution of a grid cell $H$. This is illustrated by Figure \ref{pvoutput_knmi}, where at the first location (left panels) $H$ closely follows $Y$. At the second location (right panels) this is also broadly the case, but it is also obvious that more local effects can be seen in $Y$, which are not captured by $H$. 

\begin{figure}
\centering
\includegraphics[width=0.95\linewidth]{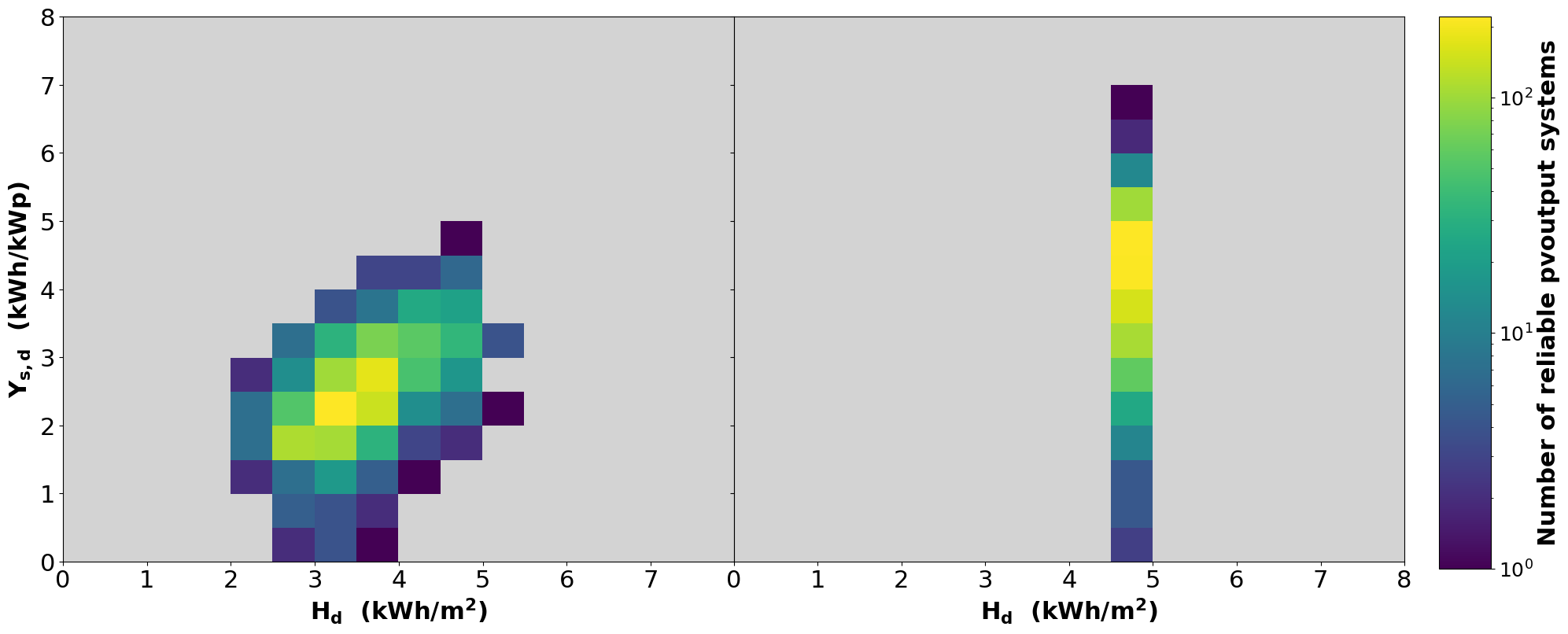}
\caption{\label{pdf_2examples}$H_{d}$ vs. $Y_{s,d}$ for all pvoutput systems on 13/06/2016 (left) and 13/09/2016 (right).} 
\end{figure}


\subsection{Determining $p_{d}(Y_{s}|H)$}
We are now in a position to evaluate $p_{d}(Y_{s}|H)$. It is trivial to compute $Y_{s,d\ i}$ and $H_{d,i}$ for each location $i$ in pvoutput. We remind the reader that $Y_{s,d\ i}=Y_{d,i}/P_{i}$. Figure \ref{pdf_2examples} shows two examples of the number density of (reliable) pvoutput systems in the $H_{d}-Y_{s,d}$ plane. The left panel (13/06/2016) shows a day with large variations in $H_{d}$ and $Y_{s,d}$, whereas the right panel (13/09/2016) shows an exceptionally clear day over the whole country, nevertheless producing a wide spread in $Y_{s,d}$, due to different efficiencies of pvoutput systems which are a function of parameters such as $\phi$ and $\theta$. We would like to remind the reader that $Y_{d}$ reflects a day in its entirety, whereas $H_{d}$ almost does. In section 2.2, we explained that $H$ are only available when the Sun's elevation is higher than $12^\circ$.


\subsection{Putting it all together}
Now we evaluate equ. \ref{equpE2} by re--writing this more intuitively, using the graphical representation of $p_{d}(Y_{s}|H)$ in Figure \ref{pdf_2examples}. We can define that for $N_{H_{d}}$ and $N_{Y_{d}}$ bins, the probabilities of $Y_{s,d}$ being observed must sum to one (equ. \ref{equPunity}). Using $p_{d}(H)$, we compute the number of systems $N_{l}$ per bin $l$, which must satisfy equ. \ref{equNd}, where $N_{d}$ is the number of systems in the database. We can use this information to compute the number of systems per bin $N_{kl}$ (equ. \ref{equNkl}). Keeping track of which systems fall in bin $l$, we can randomly draw $N_{kl}$ systems for bin $k,l$ and insert this into equ. \ref{equEd}, where $P_{m}$ is the system's power.

\begin{gather}
\sum\limits_{k=1}^{N_{Y_{d}}}\sum\limits_{l=1}^{N_{H_{d}}}p_{kl}(Y_{s,d})=1
\label{equPunity}\\
N_{d}=\sum\limits_{l=1}^{N_{I_{d}}}N_{l}
\label{equNd}\\
N_{kl} = \frac{p_{kl}}{\sum\limits_{k=1}^{N_{E_{d}}}p_{kl}}\frac{N_{l}}{N_{d}}
\label{equNkl}\\
Y_{d}=\sum\limits_{k=1}^{N_{Y_{d}}}\sum\limits_{l=1}^{N_{H_{d}}}N_{kl}Y_{s\ kl}\sum\limits_{m=1}^{N_{kl}}P_{m}
\label{equEd}
\end{gather}

\begin{figure}[!htb]
\centering
\includegraphics[width=0.95\linewidth]{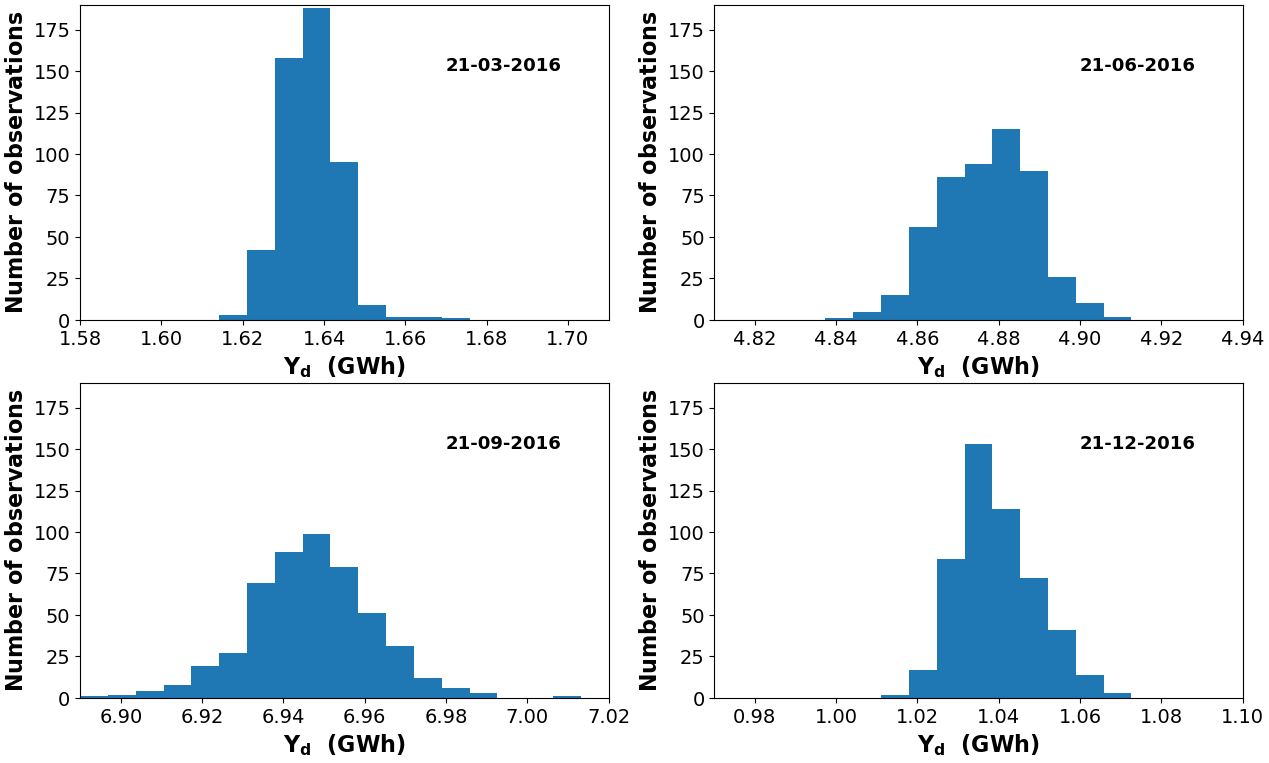}
\caption{\label{yield_2016_dailybasis}Distributions of $Y_{d}$ for all database PV systems (according to scenario 1) on the Summer and Winter solstices and the Spring and Autumn equinoxes.}
\end{figure}

\noindent We can perform Monte Carlo simulations by repeatedly evaluating equ. \ref{equEd}, resulting in a probability density function, indicating the mean ($\mu_{Y_{d}}$) and standard deviation ($\sigma_{Y_{d}}$) of our estimates for $Y_{d}$. Figure \ref{yield_2016_dailybasis} shows the density functions for four different days (Spring and Autumn equinoxes, Summer and Winter solstices). Here, we performed the simulations 500 times, allowing us to obtain smooth functions.\\

\noindent Our decision to bin the data in $0.5~kWh/m^2~\textrm{x}~0.5~kWh/kWp$ bins, as can be seen in Figure \ref{pdf_2examples} is motivated by practical concerns. We want to produce a simple, easy and intuitive model allowing us to easily read off $p_{kl}(Y_{s})$. The size of our bins is chosen in such a way that the resolution is high enough such that meaningful differences in $Y_{s}$ may be discerned, while at the same time keeping the resolution low enough, increasing the chances that each value of $H$ at a location in the database is also observed in $p_{d}(Y_{s}|H)$ from pvoutput. For those systems which have a value $H$ that is not observed in $p_{d}(Y_{s}|H)$, we use our estimate of $Y_{s,d}$ for all the present systems and multiply this by the systems' $P$. We note that the fraction of such systems is always lower than 1\%.

%% file: s5_representativeness.tex
\section{Representativeness}
Up until now our method has made one very big assumption: $p_{d}(Y|H)$ is representative for the whole PV system population. We also saw in section 3 that our data cleaning process resulted in varying populations of pvoutput systems per day (see Figure 7). Both of these elements will influence our calculations for $Y_{d}$ and $Y_{a}$. We therefore identify two refinements we can make to our model. In section 5.2 we explore what effect selecting different subpopulations has on the computation of $Y_{d}$ and hence $Y_{a}$. But before doing so, we must first define a method which allows us to re--weight $p_{d}(Y|H)$ such that the same populations are contained within each $p_{d}(Y|H)$ of the entire year. This we do in section 5.1. To make any progress on either of these refinements, we need to determine which variables characterise the population. In section 2.1 we already highlighted that $\phi$, $\theta$, $\epsilon$, $(l, b)$ of the PV systems will influence the determination of $Y_{d}$ and $Y_{a}$.

\subsection{Re--sampling pvoutput} 
New PV systems are continually placed throughout the year, with a peak in the Spring and Summer months\citep{SNsolar}. This means, at least in theory, that the distributions of system variables $\phi$, $\theta$ and $\epsilon$ can vary as a function of time. In the absence of these distributions, we argue that they can be assumed to remain constant in time. This we motivate through large number statistics: if a large population already exists with some distribution $p(x)$ and a small -- relative to the total already present -- new number of systems is continually placed, then it seems probable that these effects will average or smooth out. An obvious exception to this would be a large solar park which started generating energy from one day to the next and had a set up capable of heavily skewing $p(x)$. While our assumption would seem to hold over the course of different days, it is less obvious whether this should be the case over several years. Past research has also investigated the issue of representativeness e.g. \citet{killinger2018} derive probability distribution functions for $\phi$, $\theta$, capacity and $Y$. Unfortunately, in the case of the Netherlands, this is based largely on pvoutput data (75\% of PV systems) supplemented with other smaller data sources, therefore not helping us much.

\subsubsection{Re--sampling $\phi$, $\theta$ and $\epsilon$}
\noindent Making days consistent with each other, in terms of $x$ can be achieved by choosing a fictitious ground truth for $p(x)$ and adjusting $p_{d}(x)$ accordingly. Rather than inventing $p(x)$, we can choose it to be $p_{1}(x)$: the distribution on the first day of the year.

\begin{equation}
p_{d}(\phi,\theta,\epsilon)\sim p_{1}(\phi,\theta,\epsilon) \quad\text{where}\quad 1 < d\leq 365 \footnote{In the case of 2016 this is in fact 366 since this is a leap year. This also applies to equation \ref{equintweighting2}}
\label{equintweighting}
\end{equation}  

\noindent Integer weighting can be used to satisfy equ. \ref{equintweighting}: certain systems are randomly selected more than once, while others may be dropped. This process is repeated until equ. \ref{equintweighting} is satisfied. Our choice for integer weighting, as opposed to floating point weighting, is motivated by the fact that PV systems are discrete quantities. If we adopted floating point weights, this would produce odd situations whereby systems can be partially duplicated or discarded. To satisfy equ. \ref{equintweighting}, we can bin a pvoutput variable $x$ according to equ. \ref{obsperbin}. Then, summing all these bins for variable $x$, we satisfy equ. \ref{total}, which must always equal one. Finally $x_{i}$ and $x_{i+1}$ in equ. \ref{obsperbin} are defined by equ. \ref{binsize}, which does nothing more than defining the lower and upper limits of bin $i$ for a given bin size $\Delta x$. Finally, the number of bins is defined by equ. \ref{nbins}.

\begin{gather}
N_{bins}=\frac{x_{max}-x_{min}}{\Delta x}
\label{nbins}\\
x_{i}=\sum_{i=1}^{N_{bins}}x_{min}+i\Delta x
\label{binsize}\\
  p_{i}(x) = \frac{1}{N}\sum_{n=1}^{N}
    \begin{cases}
      x_{i}<x_{n}<x_{i+1} & 1\\
      \text{else} & 0\\
    \end{cases}   
\label{obsperbin}\\    
p_{d}(x)=\sum_{i=1}^{N_{bins}} p_{i}(x)=1
\label{total}
\end{gather}

\subsubsection{Re--sampling $H$}
\noindent Now we must account for one final variable: $p(l,b)$. With the aim of making $p_{d}(Y|H)$ as accurate as possible, the geographic number density of systems in pvoutput should match the density observed in the database. If, for example, 40\% of the pvoutput systems, used to construct $p_{d}(Y|H)$, lie in the West of the country on day $d$, while in the database this is 20\%, then our estimation of $Y_{d}$ could end up being too optimistic. \\

\noindent In practice, it's very difficult to apply integer weighting to $(l, b)$, since you would have to agree on bin sizes for $(l, b)$. In the case of $\phi$, $\theta$ and $\epsilon$ it makes sense to keep these bins constant, since these quantities do not change. For $(l, b)$ you would have to define an aggregation bin size that would make sense for that day given $H$, e.g. on a perfectly sunny day over the whole country, $(l, b)$ (e.g. 13 September in Figure \ref{pdf_2examples}) could essentially be the size of the country, whereas on a day with a lot of local weather effects, a different aggregation level would be necessary. Since the sample size of pvoutput is too small in any case to split it up into smaller portions, we can use a proxy for $(l, b)$, which is $H$ itself. We can apply integer weighting to $H$ observed in pvoutput such that its distribution satisfies the distribution in the database (equ. \ref{equintweighting2}). 

\begin{equation}
p_{d}(H)\sim D_{d}(H)  \quad\text{where}\quad 1 \leq d\leq 365 
\label{equintweighting2}
\end{equation}

\begin{table}
\caption{\label{table:minmaxdelta} Minima, maxima and bin sizes for the four different parameters: $\phi$, $\theta$, $\epsilon$ and $H$.} 
\centering
\fbox{%
\begin{tabular}{c|c c c}
\hline
parameter         & min     & max     & $\Delta$ parameter   \\ [0.5ex] 
\hline\hline
\textbf{$\phi$}          & $\ang{0}$      & $\ang{360}$    & $\ang{45}$                \\
\textbf{$\theta$}        & $\ang{0}$      & $\ang{90}$     & $\ang{15}$                \\
\textbf{$\epsilon$}      & -1             & 1              &  1                        \\
\textbf{$H$}             & min($H$)       & max($H$)   & $500~Wh/m^2$              \\ [1ex] 
\hline
\end{tabular}}
\end{table}

\noindent Table \ref{table:minmaxdelta} shows the minima, maxima and bin sizes for $\phi$, $\theta$, $\epsilon$ and $H$. It should be noted that bins for $\epsilon$ do not correspond to anything physically: 0 means $\epsilon=1$, -1 means $\epsilon>1$ and 1 that $\epsilon<1$. Our choice for the other bin sizes is motivated by practical limitations: $\Delta\phi=45^\circ$ because pvoutput only allows the input of one of eight cardinal signs. We choose $\Delta\theta=15^\circ$ such that we retain a statistically significant number ($\sim 100-200$) of PV systems per bin. We choose $\Delta H=500~Wh/m^2$ since this is what we already decided earlier on in Section 4.4 when combining $H_{d}$ and $Y_{d}$ (see Figure \ref{pdf_2examples}).\\

\begin{figure}[!htb]
\centering
\includegraphics[width=0.96\linewidth]{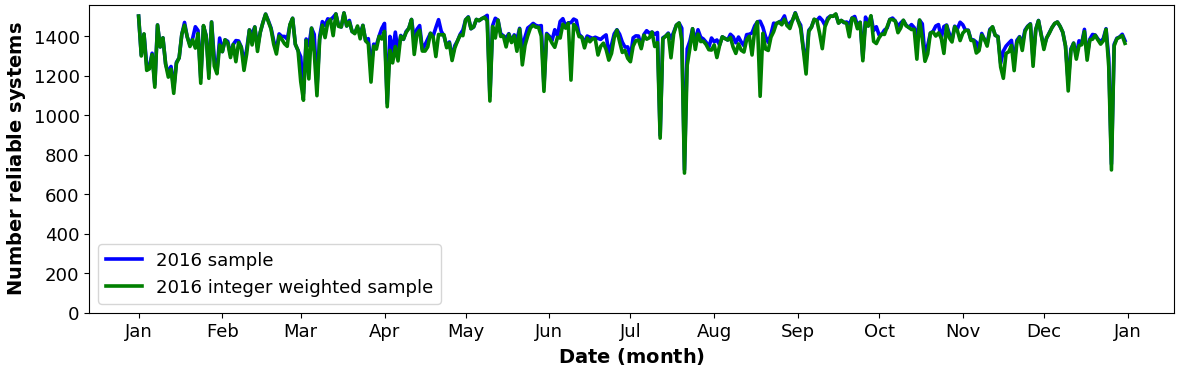}
\caption{\label{Nids_per_day_comparison}The data cleaned reliable set for 2016 (blue), along with a modified set (green).}
\end{figure}

\begin{figure}[!htb]
\centering
\includegraphics[width=0.87\linewidth]{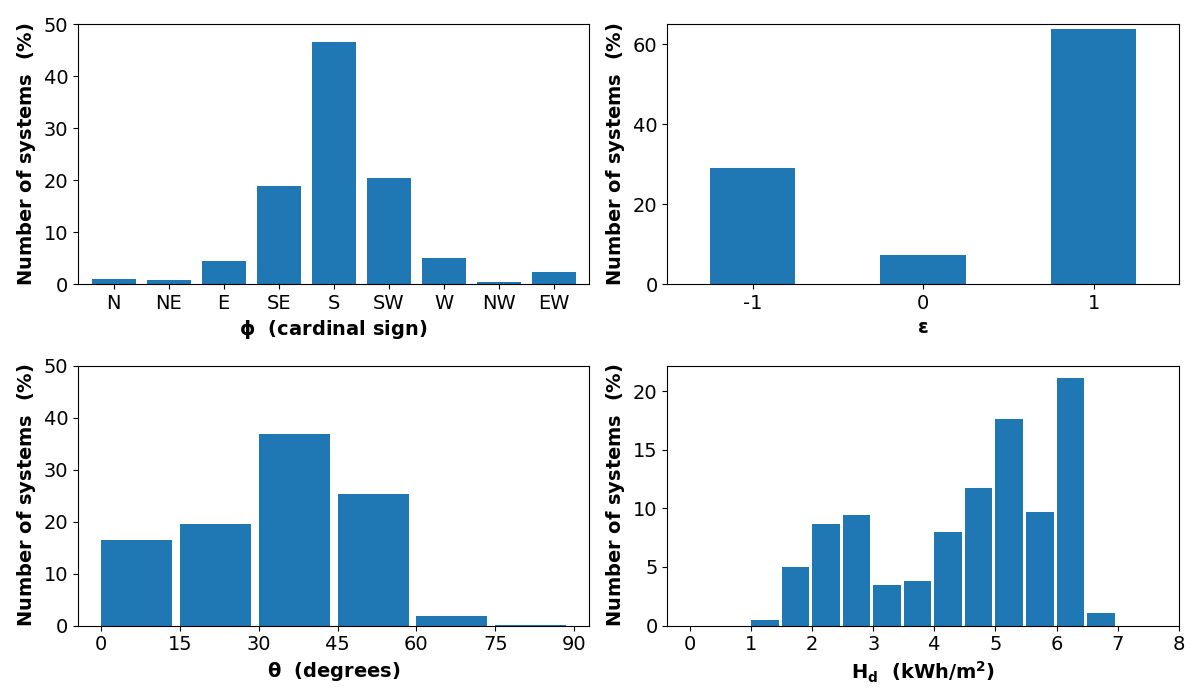}
\caption{\label{reference_dist_4D}$p_{\textrm{1/6/16}}(\phi,\theta,\epsilon)$ and $p_{\textrm{1/6/16}}(H)$, integer weighted w.r.t. 1 Jan 2016 (scenario 1.)} 
\end{figure}

\noindent We draw the reader's attention to the fact that our earlier decision for integer weighting means we cannot exactly satisfy equ. \ref{equintweighting} and \ref{equintweighting2}. This is why we allow for a leeway of 1.5\% when trying to satisfy these equations. The choice for this number is a pragmatic one: it is lenient enough to allow us to efficiently implement our procedure, but strict enough that the equations are almost exactly satisfied. Figure \ref{Nids_per_day_comparison} shows the effect of integer weighting: On some days, the overall number of systems increases while on most days the set decreases. Figure \ref{reference_dist_4D} shows $p_{\textrm{1/6/16}}(\phi,\theta,\epsilon)$ and $p_{\textrm{1/6/16}}(H)$, which have been normalised so they satisfy equ. \ref{equintweighting} and \ref{equintweighting2}.\\

\noindent As stated above, the procedure of integer weighting has the consequence that the two constraints in practice will never both be satisfied exactly. By making these constraints `softer', i.e. allowing an interval around an exact match, they become probabilistic in nature, so that it is advisable to generate multiple realisations of the distributions, all within that small allowed interval. Given that each realisation is itself a sample of between 800 and 1400 instances, a modest number of 50 realisations of the distributions is sufficient to ensure that an average over that ensemble of distributions can be used for this analysis.

\subsection{Choosing different priors}
We now return to our second refinement. In equ. \ref{equmarg} of Section 4.1, we saw that it is possible to evaluate $p(Y|H)$ as the marginal likelihood, with a prior $p(x|H)$. We can now decide to make different selections for $p(x|H)$ and propagate these through in our calculations. For example, what will $Y_{d}$ and $Y_{a}$ be if only all South--facing systems are selected? Experimenting with different choices of $p(x|H)$ will give us a sense of how much our current estimates at SN could vary depending on what the true specifics are of the Dutch PV system population. Before exploring this `scenario' testing more in Section 6.1, we summarise our methodological framework from sections 4 and 5 for the reader:

\begin{enumerate}
\item Make a choice for the prior $p(x|H)$, e.g. all systems face South.
\item Select all systems in pvoutput on 1 January that satisfy $p(x|H)$.
\item Correct $p_{d}(\phi,\theta,\epsilon)$ for day $d<1\leq 365$ and repeat this 50 times.
\item Randomly select one of the 50 `realisations' of $p(Y|x)$ and insert into $p(Y|H)$
\item Calculate $p(Y)$.
\item Repeat the last two steps 500 times
\item Estimate $\mu_{Y_{d}}$ and $\sigma_{Y_{d}}$ 
\end{enumerate}

%% file: s6_results.tex
\section{Results}
\subsection{Daily national yields}
We present the results of the various different `scenarios' -- or different choices for our prior: $p(x|H)$ -- in Table \ref{table:scenarios}. The choice for these specific scenarios is driven by their feasibility: pvoutput samples are not very large and so we are limited to those scenarios which retain enough systems. For example, determining $Y_{d}$ and $Y_{a}$ by selecting only North--facing systems ($\phi=\ang{0}$) would only leave a handful of systems. With the exception of Scenario 1, which takes $p(\phi,\theta,\epsilon)$ as a given, the other scenarios explore different choices for $\phi$ (scenarios 2--6), $\theta$ (scenarios 7--9) and $\epsilon$ (scenarios 10--13). Scenarios 14 and 15 explore combinations of all three different parameters. \\

\begin{figure}[!htb]
\centering
\includegraphics[width=0.86\linewidth]{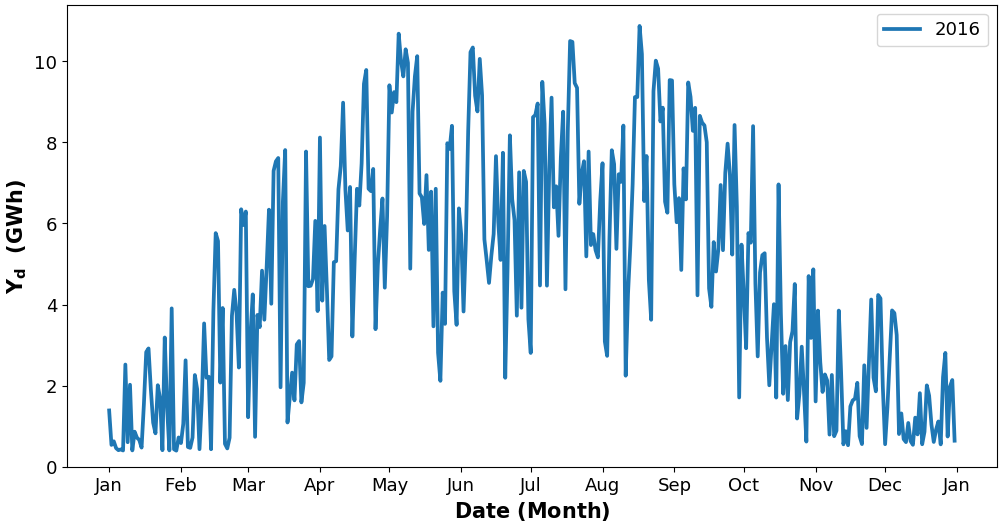}
\caption{\label{yield_2016_dailybasis_annual}$Y_{d}$ in 2016, according to scenario 2.} 
\end{figure}

\noindent Scenario 2, closely followed by 15, shows the largest $Y_{s,a}$, which is consistent with the expectation that South--facing systems outperform other set--ups. Restricting $\theta$ to a more optimal angle range (scenario 15) at the expense of relaxing $\phi$, gives similar results though. Scenario 15 outperforms scenario 14 by a small margin, suggesting that a smaller $P_{i}$ could be beneficial. Scenario 6 delivers the poorest $Y_{a}$, which is unsurprising, given that all South--facing systems are excluded. A final noteworthy mention is scenario 7: restricting PV systems to near--flat or fully flat systems, produced the second lowest $Y_{a}$, presumably due to low solar elevation angles in Winter. The $Y_{a}$ of the other scenarios lie in between these extrema. Figure \ref{yield_2016_dailybasis_annual} shows $Y_{d}$ for 2016 according to the best performing scenario 2. This visualisation nicely demonstrates what the differences are in $Y_{d}$ in Summer compared to Winter. It is worth noting that a high--irradiance Winter's day delivers relatively high $Y_{d}$ especially when comparing to a low--irradiance Summer's day.\\

\begin{table}
\caption{\label{table:scenarios}$\beta_{2016}$ and $\beta_{2017}$ (with 1$\sigma$ uncertainty margins) for 15 different scenarios.} 
\centering
\fbox{%
\begin{tabular}{c|c c c|c c c c}
\hline
Sc               & $\phi$                          &  $\theta$                     & $\epsilon$       & $Y_{s,2016}$       & $Y_{s,2017}$    & $Y_{2016}$        & $Y_{2017}$      \\ 
                 &                                 &                               &                  & ($kWh/kWp$)          & ($kWh/kWp$)       & ($GWh$)           & ($GWh$)         \\ [0.5ex] 
\hline\hline 
\textbf{1}       & $\forall$                       &  $\forall$                    &  $\forall$       & $910 \pm 0.14$       & $868 \pm 0.19$    & $1632 \pm 0.26$   & $2131 \pm 0.49$ \\    
\textbf{2}       & $=\ang{180}$                    &  $\forall$                    &  $\forall$       & $946 \pm 0.16$       & $899 \pm 0.21$    & $1697 \pm 0.30$   & $2209 \pm 0.52$ \\
\textbf{3}       & $\{\ang{135}~..~\ang{225}\}$    &  $\forall$                    &  $\forall$       & $927 \pm 0.14$       & $882 \pm 0.20$    & $1663 \pm 0.26$   & $2168 \pm 0.49$ \\ 
\textbf{4}       & $\{\ang{90}~..~\ang{180}\}$     &  $\forall$                    &  $\forall$       & $929 \pm 0.14$       & $885 \pm 0.20$    & $1668 \pm 0.26$   & $2176 \pm 0.49$ \\
\textbf{5}       & $\{\ang{180}~..~\ang{270}\}$    &  $\forall$                    &  $\forall$       & $921 \pm 0.15$       & $877 \pm 0.19$    & $1652 \pm 0.28$   & $2154 \pm 0.48$ \\ 
\textbf{6}       & $\neq \ang{180}$                &  $\forall$                    &  $\forall$       & $877 \pm 0.16$       & $838 \pm 0.21$    & $1573 \pm 0.29$   & $2059 \pm 0.51$ \\
\textbf{7}       & $\forall$                       &  $\{\ang{0}~..~\ang{30}\}$    &  $\forall$       & $895 \pm 0.18$       & $860 \pm 0.22$    & $1605 \pm 0.32$   & $2113 \pm 0.55$ \\
\textbf{8}       & $\forall$                       &  $\{\ang{30}~..~\ang{90}\}$   &  $\forall$       & $920 \pm 0.14$       & $872 \pm 0.20$    & $1652 \pm 0.26$   & $2143 \pm 0.50$ \\ 
\textbf{9}       & $\forall$                       &  $\{\ang{30}~..~\ang{45}\}$   &  $\forall$       & $923 \pm 0.16$       & $875 \pm 0.21$    & $1656 \pm 0.29$   & $2150 \pm 0.51$ \\ 
\textbf{10}      & $\forall$                       &  $\forall$                    &  $\{1\}$         & $903 \pm 0.15$       & $860 \pm 0.21$    & $1619 \pm 0.27$   & $2114 \pm 0.52$ \\
\textbf{11}      & $\forall$                       &  $\forall$                    &  $\{-1\}$        & $921 \pm 0.20$       & $875 \pm 0.25$    & $1652 \pm 0.36$   & $2150 \pm 0.62$ \\ 
\textbf{12}      & $\forall$                       &  $\forall$                    &  $\{0,1\}$       & $905 \pm 0.14$       & $864 \pm 0.20$    & $1624 \pm 0.26$   & $2122 \pm 0.48$ \\
\textbf{13}      & $\forall$                       &  $\forall$                    &  $\{-1,0\}$      & $922 \pm 0.18$       & $879 \pm 0.24$    & $1654 \pm 0.33$   & $2158 \pm 0.59$ \\  
\textbf{14}      & $\{\ang{135}~..~\ang{225}\}$    &  $\{\ang{30}~..~\ang{45}\}$   &  $\{1\}$         & $938 \pm 0.17$       & $889 \pm 0.22$    & $1684 \pm 0.31$   & $2183 \pm 0.54$ \\ 
\textbf{15}      & $\{\ang{135}~..~\ang{225}\}$    &  $\{\ang{30}~..~\ang{45}\}$   &  $\forall$       & $943 \pm 0.15$       & $893 \pm 0.21$    & $1695 \pm 0.28$   & $2195 \pm 0.51$ \\ [1ex] 
\hline
\end{tabular}}
\end{table}

\begin{figure}[!htb]
\centering
   \includegraphics[width=0.80\linewidth]{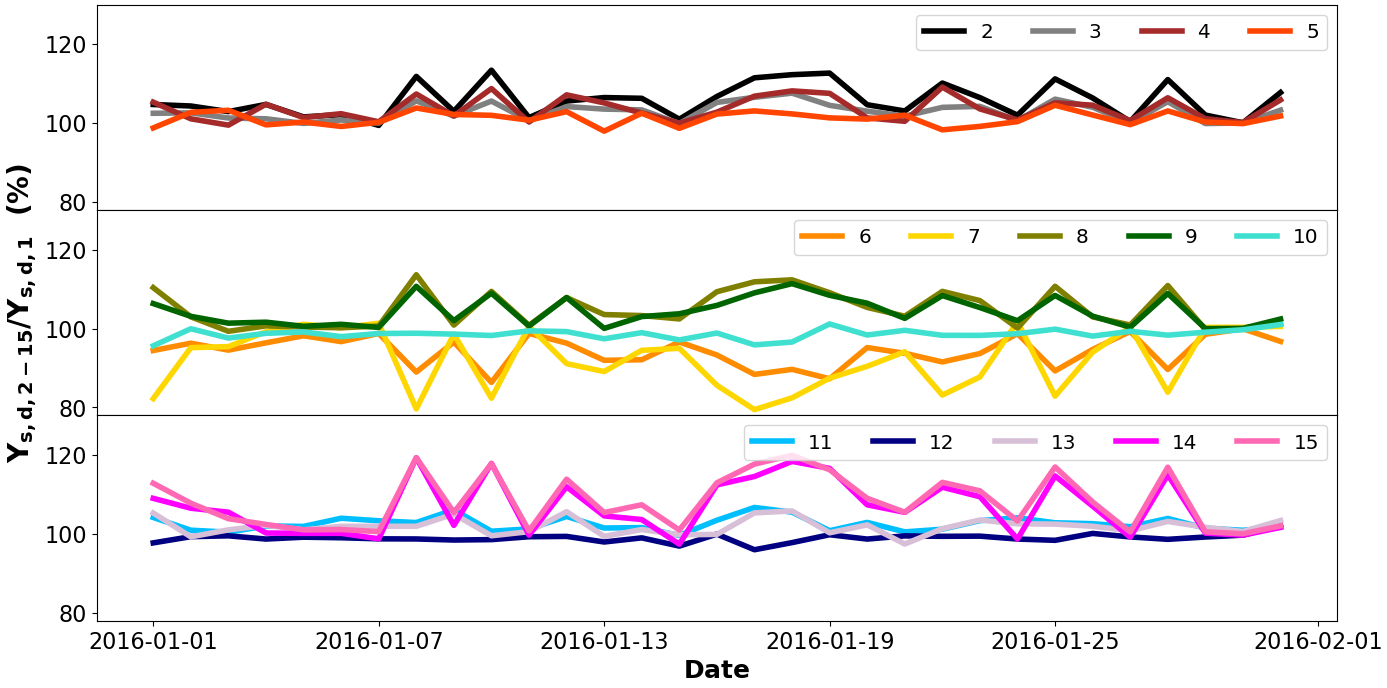}
   \includegraphics[width=0.80\linewidth]{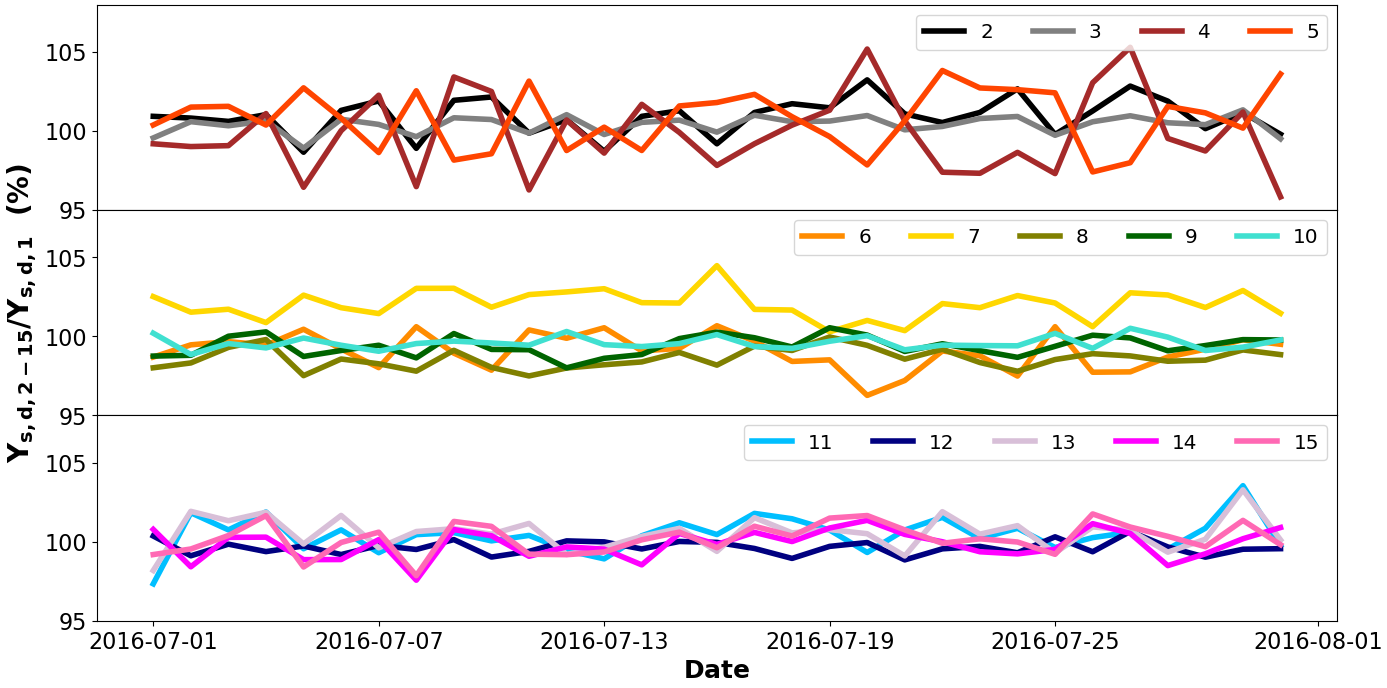}
\caption{\label{yield_2016_dailybasis_month} The ratio of $Y_{s,d, 2-15}/Y_{s,d, 1}$ for January (top) and July (bottom).} 
\end{figure}

\noindent Figure \ref{yield_2016_dailybasis_month} shows $Y_{d}$, during the months of January and July, for scenarios 2--15, normalised w.r.t. $Y_{d}$ from scenario 1. The y--axis is thus an index where a number higher than 100, means that that particular scenario led to a higher $Y_{d}$ than scenario 1 and vice versa. We include this figure because it says something about the validity of our model. We highlight one of many interesting patterns from this figure to illustrate our point. Scenario 7 can be contrasted with the other scenarios, for the two months in questions. It can be seen that this set--up leads to lower and higher $Y_{d}$ in January and July respectively, which is wholly consistent with the fact that these are low tilt systems. The less than optimal $\theta$ of these systems acts differently in the Summer with the negative effects in the Winter mitigated by a bonus in the Summer. We draw the reader's attention to the fact that for some scenarios, one can indirectly deduce the weather on that day: when contrasting scenarios 2, 3 and 4 with each other in July, one can see whether the weather was better in the morning, in the afternoon or the same. These observations support the soundness of the presented model.

\subsubsection{Discussion}
\noindent The uncertainty margins $\sigma_{Y_{s,a}}$ and $\sigma_{Y_{a}}$ are on the order of 0.05\%. We see four possible explanations for these small margins: the pvoutput sample is quite small and therefore the variation between the 50 `realisations' of the data for any choice of $p(x|H)$ could be small. Secondly, the choice of $H_{d}-Y_{d}$ bins ($0.5~kWh/m^2~\textrm{x}~0.5~kWh/kWp$) when evaluating $p_{d}(Y|H)$, could be too coarse, causing a lot of smaller scale effects to smooth out. Future work could include examining by how much $\sigma$ changes as a function of the bin resolution. Thirdly, differences on a micro level possibly average out due to the very large size of the PV systems database, which contains hundreds of thousands of entries. This is further compounded by the fact that differences in $Y_{d}$ will average out anyway when aggregating to $Y_{a}$. This is confirmed by the observation that $\sigma_{Y_{d}}$ is an order of magnitude higher than $\sigma_{Y_{a}}$. Finally, we would like to remind the reader that a bias is possible due to the choices we made when we cleaned the data. \\

\noindent Now we turn to the underlying assumptions that have been made throughout this paper w.r.t. Monte Carlo sampling. This type of sampling assumes that observations are independently and identically distributed (\textit{i.i.d.}). It is possible that on the micro level \textit{i.i.d.} is not fully satisfied due to correlations that can arise e.g. in a street with terraced houses facing the same way, it is probable that $Y_{s,d}$ is quite similar for all of the PV systems (assuming $\epsilon$ and other factors are negligible). Our method does not take these correlations into account. We would argue that the degree to which such correlations matter, depends on what aggregation levels one selects for $Y_{d}$. Since it is not our goal to present results for $Y_{d}$ on a micro--level, we argue that this effect must average out on a national level, where hundreds of thousands of database PV systems are generating energy. In this regime, $p(Y|H)$ and $p(H)$ will be the far more dominant factors when determining $\mu_{Y_{d}}$. If, for example, we are missing an important subpopulation of PV systems in $p(Y|H)$, this will have a rather larger impact. We briefly return to \textit{i.i.d.} in Section 6.2, when discussing our regional results.

\subsubsection{Comparison with Statistics Netherlands figures}
\noindent Our results for 2016 are consistently higher than those currently measured by SN: $877~\leq Y_{s,2016}\leq~946~kWh/kWp$ and $1605~\leq Y_{2016} \leq 1697~GWh$. We remind the reader that these should be contrasted with SN's estimate of $Y_{s,2016}=875~kWh/kWp$ and $Y_{2016}=1602~GWh$ \citep{SNstatline}. The result is even more significant, given that the lower range of 877 $kWh/kWp$ corresponds to an unrealistic set up: no South--facing panels (scenario 6). The picture is more mixed for 2017: $838~\leq\ Y_{s,2017}\leq~899~kWh/kWp$ and $2059~\leq Y_{2016} \leq 2209~GWh$, with $Y_{s,a}$ lying somewhere within our estimated range. 

\subsubsection{Comparison with CertiQ and SolarCare}
\noindent In section 2.1 we mentioned that SN also has $Y_{m}$ measurements for $\sim$ 1800 large PV systems (CertiQ). Unfortunately the specifics such as $\phi$, $\theta$ and $\epsilon$ are unknown. Indeed this is the reason we chose not to include the data source in our method, preferring to use it as a form of validation. Table \ref{table:certiq} shows our calculations for $Y_{m}/Y_{a}$ for pvoutput (according to scenario 1) and CertiQ. While the Winter months seem to be spot on, there are small discrepancies for the Summer months, with pvoutput showing lower $Y_{m}$, where June and September have the most striking offset. We see two possible explanations for this. Firstly, the configuration of the large PV systems may be more optimal (e.g. $\phi$), thus delivering better $Y_{m}$ in the Summer months. Secondly, panel temperatures are likely to be higher on roofs than in fields, resulting in lower conversion efficiencies (typically 5-10\% lower), and thus $Y_{m}$ \citep{drews2007}. Finally, we note that $Y_{s,2016}=904~kWh/kWp$, is in close agreement with pvoutput's scenario 1: $Y_{s,2016}=910~kWh/kWp$. SolarCare \citep{Solarmagazine} publish results for $Y_{s,a}$ and are summarised in Table \ref{table:sources}. These figures are based on some 2500 PV systems across the whole country, making their estimates very robust and worthy of comparison. We notice a small offset for 2016 and 2017 of $\sim10~kWh/kWp$ between SolarCare and our results, with our results a little more conservative. Our results are therefore broadly in line with those recorded in CertiQ and SolarCare.  

\begin{table}
\caption{\label{table:certiq}Comparison of $Y_{m}/Y_{a}$ in 2016 for scenario 1 and the large PV systems in Certiq.}
\centering
\fbox{%
\begin{tabular}{c|c c c c c c c c c c c c}    
\hline
2016 & Jan & Feb & Mar & Apr & May & Jun & Jul & Aug & Sep & Oct & Nov & Dec \\ [0.5ex]
\hline\hline
scenario 1 & 2.5 & 4.9 & 8.1 & 11.6 & 13.9 & 11.9 & 13.2 & 12.5 & 10.6 & 5.8 & 2.9 & 2.2        \\
CertiQ & 2.5 & 4.7 & 8.0 & 11.5 & 14.2 & 12.5 & 13.5 & 12.4 & 10.1 & 5.6 & 2.9 & 2.2        \\ [1ex]
\hline
\end{tabular}}
\end{table}

\begin{table}
\caption{\label{table:sources}$Y_{s,a}$ according to SolarCare \citep{Solarmagazine}, SN and our research.} 
\centering
\fbox{%
\begin{tabular}{c|c c c c}    
\hline
Year            & SolarCare         & SN           & This research   \\ [0.5ex] 
\hline\hline
2012            & 900               & 875          & n.a.              \\
2013            & 890               & 875          & n.a.              \\
2016            & 920               & 875          & 910               \\
2017            & 880               & 875          & 868               \\ [1ex]
\hline
\end{tabular}}
\end{table}

\subsubsection{Comparison between 2016 and 2017}
\begin{figure}[!htb]
\centering
\includegraphics[width=0.85\linewidth]{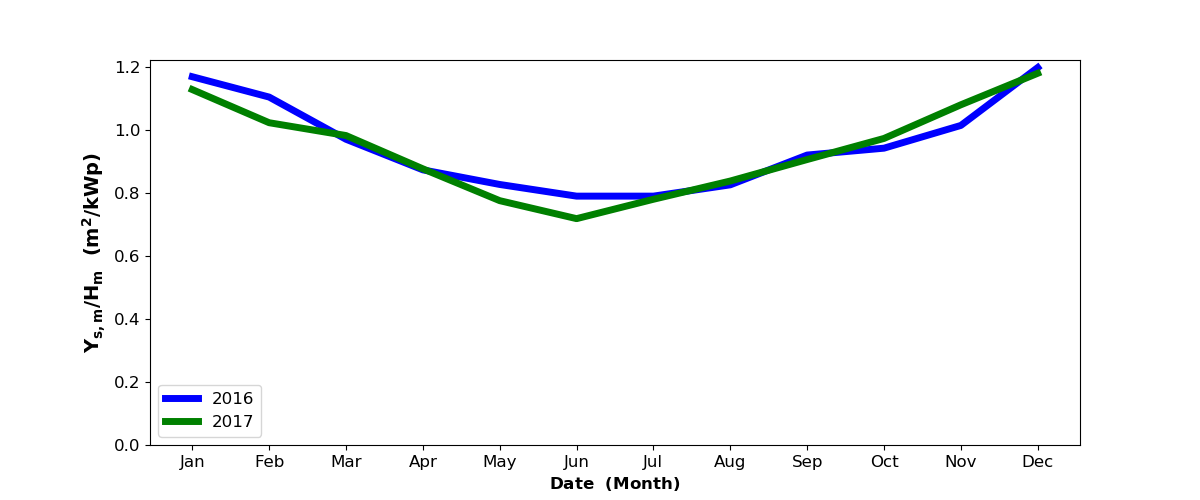}
\caption[Time series of daily produced solar power in the Netherlands]{The performance ratio ($Y_{s,m}/H_{m}$) for 2016 (blue) and 2017 (green).} 
\label{PVirr_2016_2017}
\end{figure}

\noindent Our measurement of $Y_{s,2016}$ seems to be consistent with the -- higher than 30 year average -- measurement of $H_{2016}=1039kWh/m^{2}$, as recorded by the KNMI's central weather station: De Bilt (please see Section 2.1 for more information) \citep{knmi2016_jow, knmi2012_jow, knmi2013_jow}. This cannot be said for $Y_{s,2017}$, which, despite a higher than average irradiance (1020 $kWh/m^{2}$) \citep{knmi2017_jow}, remains rather average when comparing to SN's $Y_{s}=875~kWh/kWp$. Figure \ref{PVirr_2016_2017} shows the monthly performance ratio $PR_{m}=Y_{s,m}/H_{m}$, using $H_{m}$ as was measured at De Bilt, for 2016 (blue) and 2017 (green), according to scenario 1 (see e.g. \citet{reich2012} for more information regarding the performance ratio). In this figure, it is striking that $PR_{may}$ and $PR_{jun}$ are lower in 2017 than 2016, suggesting the conversion of $H$ to $Y$ seems to have been less efficient. We note that June 2017, in particular, was an abnormally warm month with eight Summer days ($T_{max}>\ang{25}C$) and two tropical days ($T_{max}>\ang{30}C$). From the weather records, it appears that June 2017 was the warmest June month on record \cite{knmi2017june}. June 2016 contained five Summer days, consistent with the average number expected in June. From the literature it is known that panel efficiency drops with about 0.4\% per $^\circ$C. For a sunny day with high ambient temperature a panel on a roof can reach \ang{80}C, thus leading to a relative reduction of 25\% in efficiency (see e.g. Figure 5 in \citet{drews2007}). We therefore hypothesise that a possible temperature effect was responsible for the decrease in $Y_{2017}$.

\subsection{Daily regional yields}
\begin{figure}[!htb]
\includegraphics[width=0.92\linewidth]{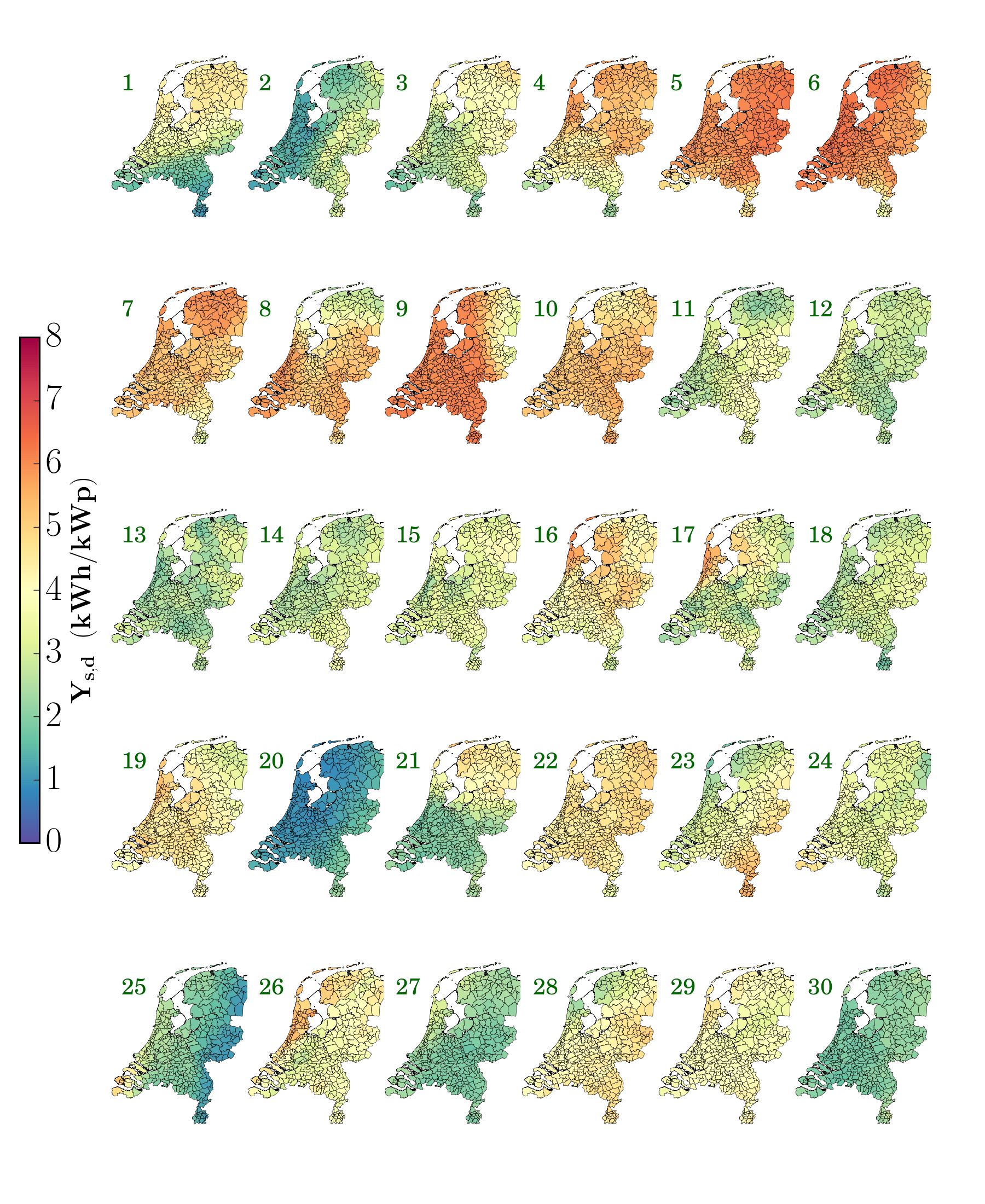}
\caption{\label{regional_june}$Y_{s,d}$ per Dutch municipality for each day in June 2016.} 
\end{figure}

\begin{figure}[!htb]
\centering
\includegraphics[width=0.48\linewidth]{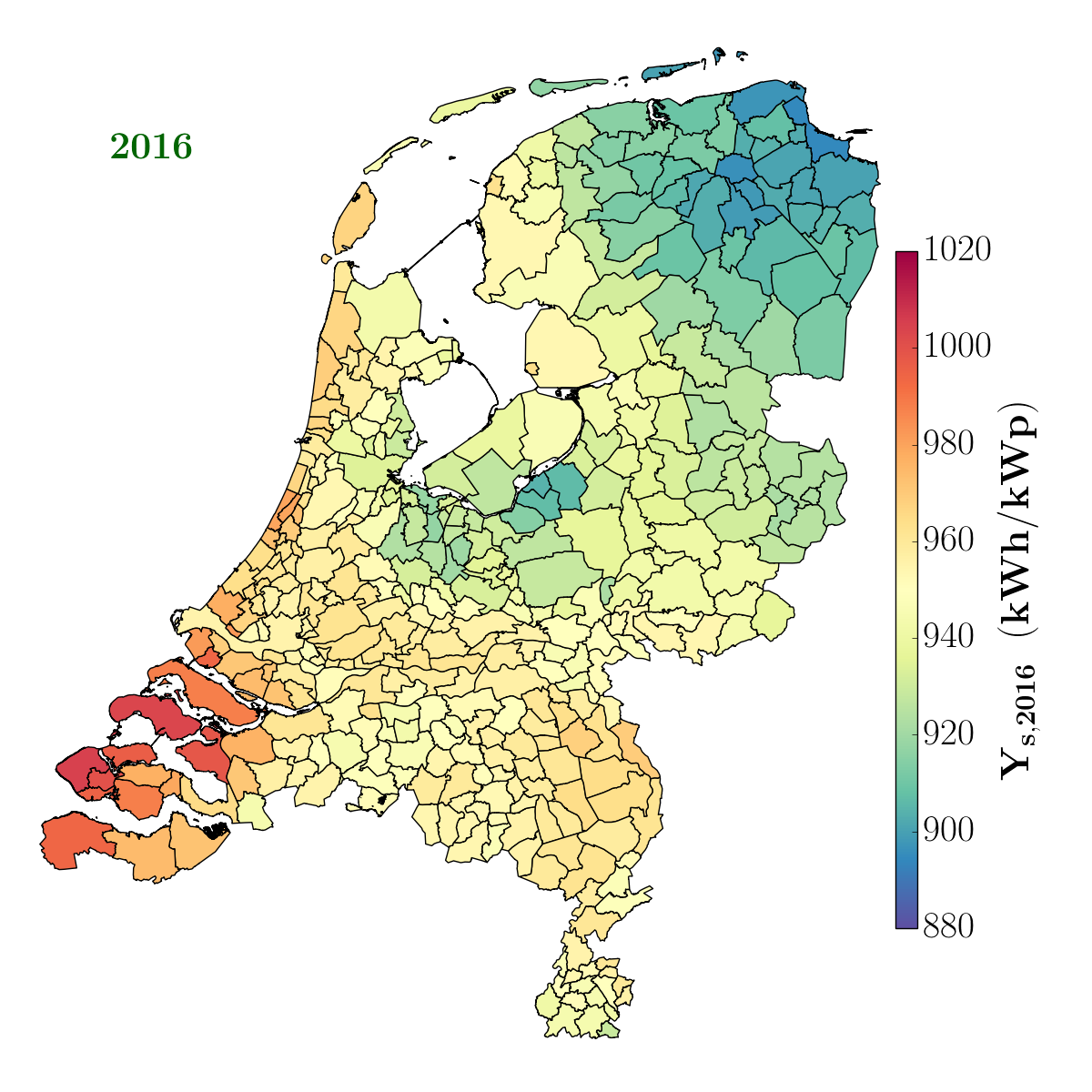}
\includegraphics[width=0.48\linewidth]{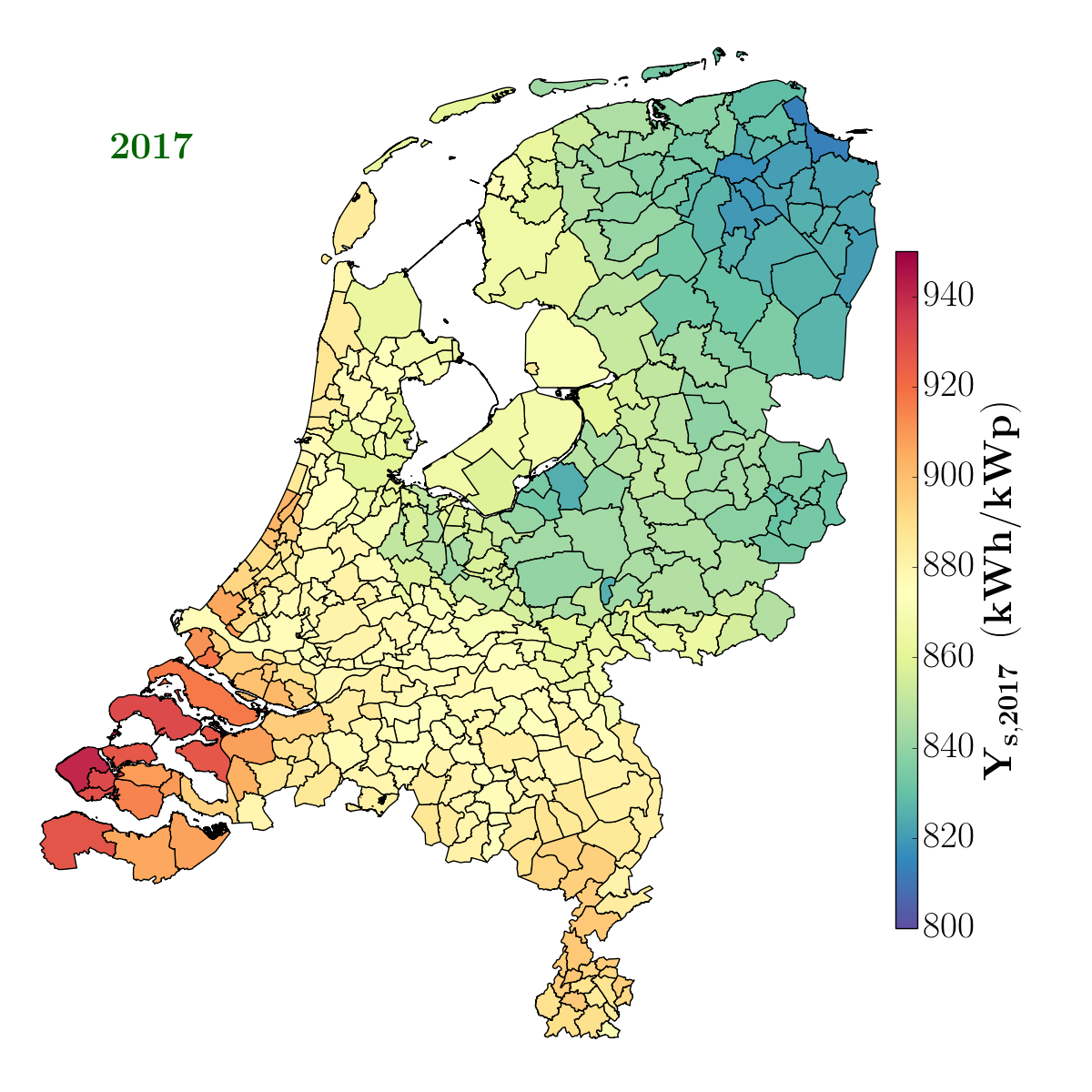}
\caption{\label{regional_annual_20162017}$Y_{s,2016}$ (left) and $Y_{s,2017}$ (right) per Dutch municipality according to scenario 2.} 
\end{figure}

\noindent Figure \ref{regional_annual_20162017} shows $Y_{s,mun}$ for 2016 and 2017, as a result of calculating equ. \ref{equbetajd} and \ref{equbetamun}. These figures validate our model because it shows patterns one expects: coastal regions enjoy more sunshine and therefore also higher $Y_{s,a}$ or $Y_{a}$. The lower $Y_{s,2017}$ compared to $Y_{s,2016}$ is also entirely consistent with other results (see section 6.1). The overall pattern of $Y_{s,a}$ looks similar between 2016 and 2017. Finally Figure \ref{regional_june} shows the $Y_{s,d,mun}$ for every day in June 2016, showing a large variation in the daily patterns.

\subsubsection{Discussion}
\noindent While our method is constructed in such a way that we produce $Y_{d}$ per database location, we would like to emphasise that this does not mean that we can accurately predict $Y_{d}$ produced at any location in the Netherlands. Rather, the point of our method is that, when aggregated to sufficiently large enough areas, we expect $Y_{d,mun}$ on those levels to come close to reflecting reality, should the PV systems be installed according to the scenario set--ups as specified in Table \ref{table:scenarios}. Even then, it should be noted that if the relative share of e.g. a large solar park is high compared to household PV systems in a given municipality, then this could influence the measurement of $Y_{d,mun}$. In light of the earlier discussion on \textit{i.i.d.}, it should be noted that re--computing the yields locally in this fashion, account to some degree for the correlations we mentioned earlier.\\ 

\noindent We remind the reader again of the fact that $H_{d}$ is only provided when the solar elevation is higher than \ang{12}. This means that the offsets that are calculated in equ. \ref{equbetajd} become more uncertain the closer we get to the Winter solstice: the Sun's highest elevation angle keeps reducing, meaning that larger portions of the day will not be captured in the totals of $H_{d}$. It is, for example, possible that $Y_{s,d\ j}\approx \mu_{Y_{s,d}}$, but that there was a lot more $H_{d,j}$ than average in the early morning and late afternoon, which would increase $Y_{d,j}$. This is not captured in our method. Finally it should be noted that the calculation of equ. \ref{equbetajd} assumes $H_{d,j}$ acts on $Y_{d,j}$ in the same way, regardless of the location. This also need not be true, since other local weather factors, especially temperature, could have an effect on $Y_{d,j}$.

%% file: s7_summary.tex
\section{Conclusion and Summary}
We have presented a new method, in the form of a classical estimation problem, to determine the daily, national and regional yield generated by photovoltaic systems in the Netherlands. We combined information from two data sources to achieve this: pvoutput, an online portal with measurements of solar energy yield and high resolution irradiance data from the Royal Netherlands Meteorological Institute. We generate daily probabilistic functions indicating the most likely specific daily yield given the irradiance. These functions are applied to our database containing almost all PV systems in the Netherlands, producing daily and national energy yield estimates, whose uncertainties are estimated through Monte Carlo sampling. The national estimates are converted to regional ones, by comparing the local with the mean irradiance observed.\\

\noindent The effect of choosing different priors, relating to the distributions of orientation, tilt and inverter to PV capacity ratio of the systems, is explored for the daily and hence annual yield. For 2016, we find specific annual yields in the range of 877--946 $kWh/kWp$, which is consistently higher than 875 $kWh/kWp$ used in the current method. For 2017, Statistics Netherlands may have overestimated the yield, since we find specific yields in the range of 838--899 $kWh/kWp$. These results highlight the need for a specific yield to be determined on a daily and annual basis which is a function of the irradiance. We believe the results of our new method can be of great use for policy making at a municipality and provincial level, where efforts are undertaken to stimulate renewable energy initiatives.

%% file: acknowledgements.tex
\section*{Acknowledgements}
We thank Alex Priem, Dick Windmeijer, Jurri\"en Vroom, Reinoud Segers, Anne Miek Kremer, Andr\'e Meurink, Otto Swertz, Lyana Curier, Sofie De Broe, Michael Maseda and Nicolas Martin for their help and discussions. The authors also wish to thank colleagues at other European statistical offices which provided information on the calculation of the solar yield in their respective countries. W.G.J.H.M. van Sark acknowledges financial support from the Ministry of Economic Affairs and Climate Policy through Topsector Energy funding for the TKI-Urban Energy project PV Observatory. Finally, B.P.M. Laevens thanks his colleagues at the ministry of Economic Affairs and Climate Policy for their support.